# Intrinsic Ferromagnetism in Eu doped ZnO

M.H.N. Assadi[a], Y.B. Zhang[a, 1], M. Ionescu[b], P. Photongkam[a], and S. Li[a]

[a] *School of Materials Science and Engineering, The University of New South Wales, NSW 2052, Australia*
[b] *Australian Nuclear Science and Technology Organization, Sydney, NSW 2234, Australia*

We report room temperature ferromagnetism in as-implanted Eu doped ZnO (ZnO:Eu). To address the origin of ferromagnetism *ab initio* calculations of ZnO:Eu system are performed. Results show that the ferromagnetism is induced by ZnO point defects as Eu ions in perfect ZnO tend to align antiferromagnetically.

## 1. Introduction

Diluted magnetic semiconductors (DMSs) have attracted enormous interests because of their potential for innovative spintronics application [1]. In order to achieve Curie temperatures ($T_C$) above room temperature, DMSs are normally fabricated by incorporating transition metal or rare earth ions into a nonmagnetic semiconductor host lattice [2]. Among many candidates, ZnO-based DMSs with unique technological applications have recently been in the focus of an intense and ongoing research [3]. The net magnetization in DMS materials should not arise from ferromagnetic inclusions (secondary phases), but from localized magnetic moments of separated ions, being distributed uniformly in the host and ferromagnetically aligned via an indirect of magnetic coupling. Due to the highest magnetic moment that can be born on a single ion, Eu is considered as magnetic dopants in ZnO to achieve a high magnetization DMS. However, the solubility of Eu in ZnO is strongly limited by the lattice distortion associated with the much larger ionic radius of the rare earth atoms. The magnetic interactions between the localized impurity atoms as well as lattice distortions caused by these atoms play an important role in determining atomic arrangements in such materials. In this paper, we report on ferromagnetic properties at room temperature of Eu-doped ZnO epitaxial thin films using ion beam techniques. Then by using theoretical *ab initio* techniques, we reveal the arrangement of Eu dopants in the host lattice and relate the ferromagnetic interaction among Eu ions to ZnO's native defects.

## 2. Sample preparation

Epitaxial ZnO (0001) thin films for ion implantation were supplied by Nanovation Inc. France. The films were grown on 10 × 10× 0.5mm c-$Al_2O_3$ substrates by Pulse laser deposition (PLD). Each film has approximately 100 nm thickness. The ion implantation was performed at room temperature by using a metal vapour vacuum arc (MEVVA) ion source under vacuum, $2 \times 10^{-6}$ mbar. The Europium (Eu) ions were implanted at 45 kV with ion beam current 30 mA. Two Eu-implanted ZnO samples have been prepared at different implantation time. The amount of Eu in these two samples was characterized by Rutherford backscattering experiment was $1.253 \times 10^{16}$ and $1.670 \times 10^{16}$ atom/$cm^2$. These values are equivalent to 3% and 4% by mole respectively. The magnetic properties of Eu-doped were studied by Quantum Design MPMS SQUID magnetometer.

---

[1] Author to whom correspondence should be addressed; electronic mail: y.zhang@unsw.edu.au

## 3. Results

Fig. 1 shows isothermal magnetization curves of as-implanted $Zn_{0.97}Eu_{0.03}O$ and $Zn_{0.96}Eu_{0.04}O$ films as a function of an external magnetic field applied parallel to the film surface measured at room temperature. It demonstrates that both of films possess ferromagnetism at room temperature with coercive fields of 50 Oe and 1.81 and 2.23$\mu_B$/Eu for $Zn_{0.97}Eu_{0.03}O$ and $Zn_{0.96}Eu_{0.04}O$ respectively. This result indicates slight change in magnetization per Eu ion as Eu concentration increases. Ferromagnetism found in these samples can not be caused by Eu metal clusters or oxide forms such as. EuO and $Eu_2O_3$ since XRD results show no secondary phases.

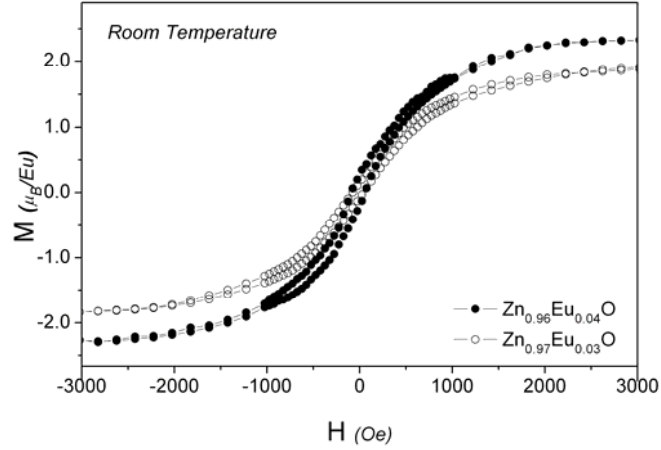

Fig. 1. M-H curve of $Zn_{0.97}Eu_{0.03}O$ and $Zn_{0.96}Eu_{0.04}O$ at 300K measured by SQUIDS magnetometer, after the subtraction of diamagnetic contribution of Al2O3 substrate. The external applied field was applied parallel to the film surface

## 4. Theoretical investigation

In our theoretical study, total energy ($E^t$) calculations were performed using plane-wave pseudopotential approach of density functional theory as implemented in CASTEP code [4] within the framework of generalized-gradient approximation. Ultrasoft pseudopotentials were represented in reciprocal space and Eu's 4$f$ were treated as valence electrons. An energy cut-off of 800 eV for plane-wave basis set and a 3×3×1 Γ-centered $k$-point grid for integration over reciprocal space were used. To model the doped ZnO, a 32-atomic 2$a$×2$a$×2$c$ ZnO supercell was adopted for calculations with two Eu ions substituting Zn sites. This was necessary for calculations of the relative energies of ferromagnetic (FM) and antiferromagnetic (AFM) spin alignments. The difference between these two energies per Eu ion is defined as $\Delta E = (E^t_{AFM} - E^t_{FM})/2$, which indicates the ferromagnetic phase stability.

In order to investigate the aggregation tendency among Eu ions, two spatial arrangements were studied, Configuration 1 in which Eu ions were separated by only one oxygen ion and Configuration 2 in which Eu ions were separated by a chain of -O-Zn-O- ions as shown in Fig. 2(a) and Fig. 2(c) respectively.

For both of configurations geometry relaxation was performed allowing internal coordinates to relax until the Cartesian components of atomic forces acting on all ions in the supercell was smaller than 0.05 eVÅ$^{-1}$ and simultaneously the energy converged to 10$^{-5}$ eV per step, per atom. The relaxed structures for both configurations are represented in Fig. 2(b) and Fig.2(d) respectively. $E^t$, $\Delta E$ and the Eu-Eu distance ($D_{Eu-Eu}$) are presented for both configurations in Table 1.

$E^t$ for Configuration 2 is lower by 44 meV than that for Configuration 1, indicating that Eu ions in ZnO do not tend do aggregate via an oxygen ion. This lowers the chance for nono-scale aggregation or secondary phase formation.

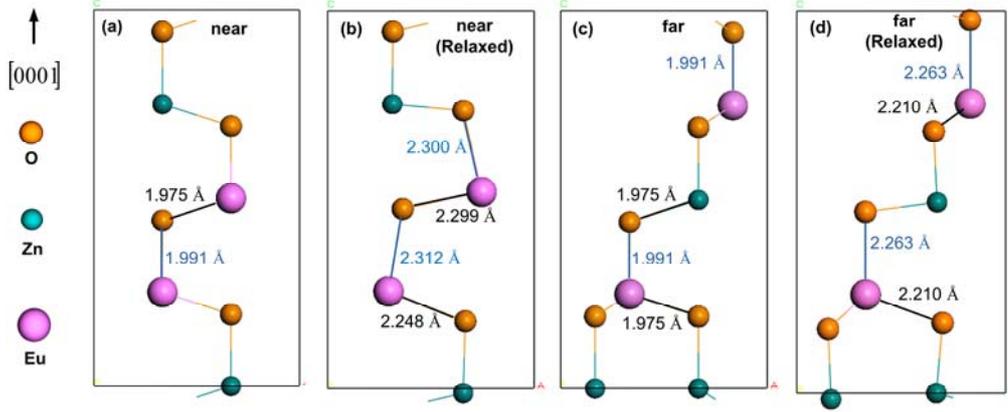

Fig.2. Schematic representation of the ZnO wurtzite structure, with Eu ions incorporated into the host lattice in Configuration 1, unrelaxed (a) and relaxed (b) and Configuration 2, unrelaxed (c) and relaxed (d). Note the magnitude of the relaxation, particularly of Eu-O bond length. The magnitude of the relaxation in Eu-O bond has decreased significantly as Eu-Eu separation has increased.

Due to Eu's larger radius, it is expected that incorporation of Eu ions in ZnO introduces local lattice distortion. Such lattice distortion is examined by analysing the difference between the Eu-O bond length in unrelaxed and relaxed structures of both configurations. According to Fig. 2(a) the Eu-O bond length in unrelaxed structure which is identical to Zn-O bond length in un-doped ZnO, is 1.991 Å along $c$ direction and 1.975 Å within $ab$ plane. In Configuration 1, after relaxation of ionic coordinates the Eu-O bond length expanded dramatically to 2.300 Å and 2.312 Å (~ 14%) along $c$ direction and 2.248 Å and 2.299 Å (~ 14%) within $ab$ plane.

As the separation of Eu ions increases in configuration 2, a slight decrease in the magnitude of the Eu-O bond length expansion is observed. Its Eu-O bond length is expended to 2.363 Å (12%) along c direction and 2.210 (11%). This indicates the expansion strain caused by substitutional Eu ions reduces as their separation increase.

Table 1. $E^t$, $\Delta E$ and $D_{Eu-Eu}$ are presented for the ZnO:Eu system. $E^t$ and $D_{Eu-Eu}$ are shown for the relaxed AFM structure. $E^t$ is presented with respect to an arbitrary origin which puts the highest calculated energy at zero.

| Configuration | $E^t$(meV) | $\Delta E$(meV) | $D_{Eu-Eu}$(Å) |
|---|---|---|---|
| 1 (near) | 0 | -20 | 3.753 |
| 2 (far) | -44 | -40 | 4.460 |

Magnetically $\Delta E$ is -20 meV for Configuration 1 and -40 meV for Configuration 2. Negative values for $\Delta E$ in both configurations suggest an antiferromagnetic interaction between Eu ions separated either by a single oxygen ion or the chain of -O-Zn-O- ions at 0 K. Since the energy associated with thermal electronic fluctuations in a given temperature is in the range of ~ $k_BT$, approximately 25.8 meV for $T$ = 300 K, these values of $\Delta E$ do not guarantee any effective coupling at room temperature. Therefore, paramagnetic behaviour is predicted at room the ZnO:Eu system.

In the next step, simulation was repeated on defective ZnO:Eu systems. Two different defects were considered: (a) oxygen vacancy ($V_O$) and (b) interstitial Zn ($Zn_I$). They were introduced into Configuration 2. $\Delta E$ for the system with $V_O$ was -5 meV which excludes the possibility of $V_O$ mediating the observed ferromagnetic interaction in the ZnO:Eu system. In the defective system with $Zn_I$, $\Delta E$ rises to +152.2 meV, which proves that the ferromagnetic state is much more stable than antiferromagnetic state in the presence of $Zn_I$. In this system the magnitude of $\Delta E$ is almost 5 times larger than $k_BT$ at room temperature which guarantees $T_C$ above 300 K for this system.

## 5  Conclusion

Room temperature magnetism was observed in for $Zn_{0.97}Eu_{0.03}O$ and $Zn_{0.96}Eu_{0.04}O$ systems. In search for the origin of observed magnetism, theoretical investigation revealed: (1) Eu ions do not tend to aggregate via oxygen, which reduces the chance of the formation of nano-scale magnetic secondary phase. (2) Ferromagnetic interaction is mediated by defects such as $Zn_I$. However $V_O$ does not stabilizes the ferromagnetic phase.

**Acknowledgments**

This work was supported by Australian Research Council (Grant Nos. DP0770424 and DP0988687) and Australian Institute of Nuclear Science and Engineering (AINSE award, AINGRA09118).